# Macroquantum effects in condensed matter

*Prof. A.M. Ilyanok*

*Abstract*

*Quantum relativistic probability description of continuous world is re-examined. The new conception of physical world is offered. It is based on deterministic description of elementary particles, on conception*

*The standard probability quantum relativistic picture of world is reviewed. The new picture of world is proposed. It is based on deterministic approach to elementary particles, discrete model of space determined by segmentation of particles and their fields and also by introduction of 9 Lorentz groups of transformations between systems of reference.*

*It is supposed that elementary particles are segmented in space and change their form dependent on phase and structural state of substance.*

*This approach allow to use linear differential second order equations to describe interactions. Maxwell equations are the partial case of them.*

*This approach allow to eliminate basic paradoxes of quantum field theory and quantum mechanics and gives a new interpretation of well known experiments.*

*To confirm the theory proposed basic parameters of macroquantum effects in condensed medium including cosmic objects are calculated. Particularly, parameters of high temperature superconductors being stable till temperature $93,5^0 C$ are calculated. It is supposed that electrons in atomic hydrogen at $6000^0 C$ at the Sun can occur in superconducting state.*

In 20$^{th}$ century classical mechanics had not being able to describe a number of experimental effects found with elementary particles. So modification of classical mechanics by attributing wave properties to particles taking simultaneously in account their relativistic properties became evident close under speed of light had been occurred. The space properly in which particle movements take place was considered as Euclidean continuous three-dimensional linear space or as four-dimensional Minkowski space. For cosmic objects it is considered as non-linear four-dimensional pseudo-Riemanian space proposed by Einstein. But such quantum relativistic concept of continuous world formed in the beginning of the century did not consider macroscopic quantum effects moreover in space.

Two most important discoveries of 20$^{th}$ century, macroquantum effects in condensed matter, was not inscribed in the concept of world being formed in such a way. They are superconductivity (G. Kamerling-Onness, 1911) and superfluidity of liquid helium (P.L. Kapitsa, 1938). But theoretical description of these effects had become exclusively complicated. Thus, theory of superfluidity created by L. Landau in 1941 for liquid $^4$HeII gives the result for theoretical calculation of critical speed of superfluid phase in 100 (!) times lower than the experimental one. It took place because Landau theory did not take into account quantum effects and, therefore, the Planck constant [1].

In theory of superconductivity coupled electrons, so called, Kooper pairs, are interacting between each other instantaneously (there are no retarding quantities in equations) [2]. It is obvious that such assumptions lead to paradoxes and violation of energy conservation law because exchange forces tend to the infinity and this fact is of no sense.

In result, one should use empirical coefficients to fit to the experiment the theories of superconductivity and superfluidity. This fact lead to low value of them for creation new materials and devises based on macroquantum effects. A large number of efforts to improve the theories using existing formulation of quantum mechanics and quantum field theory did not give positive results. It is because that quantum mechanics and quantum field theory use mathematical models not able to be realized physically. They are based on the concept of "probability wave" by which a particle is distributed in the space. And at low speed of its movement the probability to find it will be essential in volume exceeding the volume of the Earth. It is of no sense. Besides this the process of emission or absorption of a particle is changed by formal mathematical birth-annihilation operators [3,4]. These simplified models non-adequate to the physical reality do not enable to describe many effects in micro and macro world. Essentially, they have become as obstacle in developing physics in 20$^{th}$ century.

Physics always develop in alternative ways. But in the 20$^{th}$ century the new effective mechanism of stimulation of research by Nobel price awarding was formed. It was the most successful for experimental physics. But in theoretical physics it was lead to suppression of alternative approaches to the conception of world and only quantum relativistic approach was put in dogma. In result, the religious approach had been formed in the science.

We stand on the point of view that in every substance and even phase state of the substance there is special speed limit of interactions penetrating. So we suppose to consider all motions within the media to be covariant under the Lorentz boost transformations specific to it with factor $[1-(v/c_n)^2]^{-1/2}$, where $c_n = \boldsymbol{b}\boldsymbol{a}^n c$, $n$ = 0, ±1, ±2, ±3, ±4, where $\boldsymbol{b}$ is a factor connected with the properties of medium and its phase state, $\boldsymbol{a}$ is fine structure constant, $c$ is usual vacuum speed of light. The speed limits $c_n$ form the series of critical speeds of substance movement. . When $n$ = 0 one obtains usual Lorentz transformation applied only to the movement of elementary particles in vacuum. When $n$ = 1 one obtains the speed limit of electrons in metal and also the speed limit of movement of substance in empty space. Exceeding the speed $\boldsymbol{a}c$ condensed matter becomes unstable and decays what was found in accelerators of the micro drops of deuterium and from data of observation of movement of cosmic objects in galaxies [5]. When $n$ = 2 and $\boldsymbol{b} = 3c_2$ coincides with average orbit speed of Mercury. Besides this condition $n$ = 2 gives for $c_2$ the critical speed of electrons in low-temperature superconductors. Condition $n$ = 3 yields the critical speed of electrons connected with phonons in low-temperature superconductors. When $n$ = 4 the critical speed of superfluid vortexes in liquid Helium is achieved. Condition $n$ = −1 leads to the speed of internuclear interactions along crystal planes in the Moessbauer effect. Value $n$ = −2 gives the minimum speed of gravitational waves in the Sun system. When $n$ = −3 one obtains the speed of gravitational interactions within the core of the Galaxy. If $n$ = − 4 the speed limit of any gravitational interactions is achieved. For other values of $n$ we did not find appropriate physical phenomena. From such point of view elementary particles can not exceed the speed of light as it was stated in special theory of relativity. But if one considers the whole set of possible Lorentz groups it is possible to describe transferring information with the speed much more than the speed of light if $n$ = −1, −2, −3, −4.

The second serious problem in physics is the problem what is the dimension of space and what is the space structure of fields. To consider the problem usually the many dimensional approach is wide use. Let us follow not very common way of decreasing of a number of dimensions. In this way two-dimensional deterministic objects, vortex rings, making their any possible space combinations, can represent the three-dimensional world. In this way the concepts of atom and electron can be formulated in words of the theory of knots as a locking of $N_a$ knots. Let us remind that a knot is a smooth insertion of union of finite number of non-crossing circles in three-dimensional space $\boldsymbol{R}^3$. All possible combinations of vortex rings in the space give us a feeling of thre-dimensionality of the world.

Turning to the history of physics one can find that these ideas are not discovered at the first time. In mathematics, E. Cartan introduced fibre bundles of the space connected with world time. The geometry of every spatial cross-section was absolutely flat [6]. Earlier, in 1867 lord Kelvin supposed atoms as knotted vortex tubes of the ether like electromagnetic vortex tubes of Maxwell. This idea had been developing by his followers and it the following arguments were put onto the base of it [7]:

- Stability. Stability of substance might be explained by the stability of knots (the stability of their topological type).
- Diversity of chemical elements. It might be explained by the diversity of non-equivalent knots.
- Spectra. Spectral lines of atoms might be explained as oscillations of correspondent vortex tubes
- Transmutation. In up-to-date words it is an opportunity for elements to transmute one into other under high energies of interaction might be connected with re-combining of knots using cuts of knots.

Developing such deterministic approach to the micro world one can interpret the well known experiments from the new point of view.

Let us suppose that some constant $N_a$ should be integer number the closest value of which one may find from the measurement of fine splitting in spectrum of atomic Hydrogen:

$$N_a = \frac{2\pi}{\alpha}\sqrt{1-\alpha^2} = 861, \quad (1)$$

where α is the value involving spin and relativistic effects of electron movement in the atom. It is called as fine structure constant. It defines the cross-sections of electromagnetic interactions. The value of it follows from (1) and equal to $\alpha^{-1} = 137{,}0360547255...$ that coincides with the experimental results till the 7$^{th}$ significant digit. From (1) it follows that α is geometrical non-dimensional characteristic of the space identically connected with $N_a$ and π.

Analysis of energy spectra of stable movement of microparticles and new laws of quantisation of energies and orbits of cosmic objects [5], the supposition that the space is discrete not only in micro world but in the macro world follows. Evidently, the single way to explain that the space is discrete is to close the one-dimensional space – thin thread – round two-dimensional space – the plane. In result the thin ring is appeared.

Let us suppose the electron at the minimum of its energy as the thin uniformly charged ring with the charge $e$ rotating round its axis with the speed $\alpha^2 c$. Cross-section of interaction of such electrons is minimized. Such state of the electron one can observe in vacuum when it moves with the speed relative to the laboratory reference system less than $\alpha^2 c$ and in superconductors [8]. Diameter of the such electron may be find from the experiment when tunneling the electron through vacuum gap takes place. It if found experimentally that the tunnel effect disappears when the distance between electrodes is about 8 nm [9]. This exclusively important experimental fact is constantly ignored. Let us assume that the radius of such ring electron is connected with world constants [8]:

$$r_0 = \hbar/(m_e \alpha^2 c) = 7{,}2517 \text{ nm}. \quad (2)$$

When the speed of electron becomes more than $\alpha^2 c$, the ring electron will transform into the new the single stable state in the medium with self-interaction [10]. It will roll up in uniformly charged shallow tore of smaller size rotating round its axis with the speed $c$. In this state the electron will be three-dimensional object as it is shown on the fig. 1.

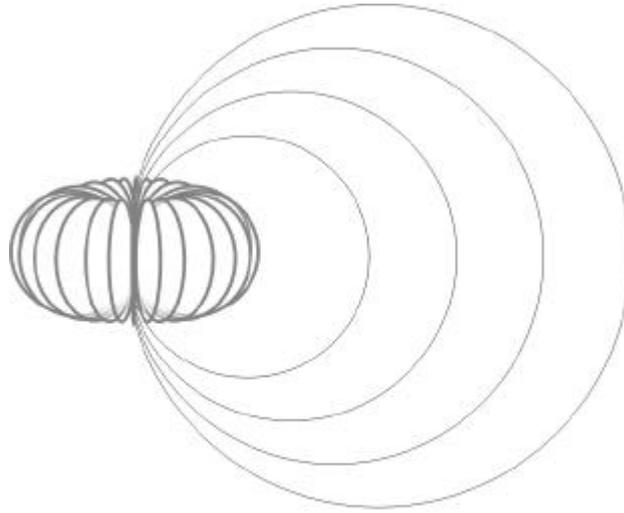

Figure 1. Electron model
Its electromagnetic field for one cross-section

The size of the tore is such that closed geodesic line on it will roll strictly 861 times round the surface of the tore. The large radius of the tore is equal to $r_l = a^2 r_0 = \hbar/(m_e c)$. This value is found experimentally from the Compton scattering of gamma-quanta on electrons. The lowest radius of the tore will be equal to $(1-a)r_l$. So the electrostatic potential hole with radius $ar_l$. Its size totally coincides with the classical electron radius.

It is found experimentally that scattering of electrons on each other or the scattering phonons on electrons in three-dimensional space is happening just at such distances. The model proposed enables us to avoid three basic paradoxes of classical electron. In the classical model all mass of the electron is concentrated in electromagnetic field and the speed of the charge created its own electromagnetic field is equal to $137.036\,c$. Besides this, eigen-proper angular momentum of the electron, the spin, can not be equal to $\pm\hbar/2$ in principal. In the model presented the internal charge is moving along the tore surface with the speed $c$ but the electromagnetic mass will be of 1/137,036 part of its total mass $m_e$, and the spin of electron is describing by the eigen-proper angular momentum of the movement of the hollow tore whet it rotates round its axis and equal $\pm\hbar/2$. This fact automatically eliminates the existing discrepancies in the model of classical electron. While moving in an empty space the spin of electron is rotating along or opposite to the direction of the movement and its electromagnetic field rotates along or opposite to the direction of movement of the electron.

The electron can change its form when transition of it from vacuum to the substance is happened. Under transition of the electron from the vacuum to substance it can change its form. The most simple variant when the electron is coupling with the proton forming the Hydrogen atom. We consider this formation as proton gets into potential hole of tore-type electron. In this case the large tore radius becomes equal to $a\,r_0$, but its speed is decreasing till $a\,c$. These parameters coincide with the model of Bohr atom. And basic parameters of the electron are change only on the amount of binding energy of it with proton $m_e(a\,c)^2/2$. This concept of atom allows totally to avoid the basic paradox of movement of a charge particle round a nucleus when a classical charged particle must emit radiation while running round and, therefore, to drop down to nucleus in very short time. In the model proposed above an electron is not affected by orbital accelerations, so it can not emit electromagnetic waves. Moreover, optical spectra are easily described by discrete change of size and speed of the shell of the tore simulating an electron. Fine splitting if spectral lines are connected with segmentation of an electron and superfine one corresponds to vibration modes of the shell of the electron tore coursed by the nucleus. In classical quantum field theory these effects are ascribed to an interaction of an electron with so called "physical vacuum". In the model proposed it is not necessary to use a new theory of the ether. If an electron is in condensed substance its spin characteristics can be changed because of changing the moment of inertia because of changing of its matter distribution along the cross-section of the tore. Mass of a free electron is distributed along the shell. If the speed of rotation of the shell is less or equal to $ac$, that is always takes place in condensed matter, so the electron mass is uniformly distributed along the cross-section of the tore. In this case its moment of inertia is changed and its spin will be equal to $\pm(\sqrt{3}/2)\hbar$. Such 'fractional' spin electrons have in paramagnetic materials.

Let us now consider the spatial structure of the external fields of an electron. Considering an electron as a tore let us cross it 861 times by transversal plane along its large diameter like lists of a book opened on 360°. So fields of rings or discs appeared by this section form its electrostatic field. Closed lines in their planes represent magnetic field of the electron like co-central circles on each plane of the ring or disc. The configuration of the field of an electron is presented on the fig.1. The $N_a$ defines a number of transversal layers but $n$ in $a^n$ defines a number of longitudinal layers. Interaction between an electron and external electromagnetic field is because of transversal layers. Taking into account such character of interaction and 'fractional' spin one may calculate anomalous magnetic moment of an electron what gives

$$\boldsymbol{m} = (1 + N_a^{-1} - 4/3\, N_a^{-2})\boldsymbol{m}_B = 1.0011596415\ 95\ \boldsymbol{m}_B, \tag{3}$$

where $\boldsymbol{m}_B$ is the Bohr magneton. It coincides with experimental value till the seventh digital after the dot.

To calculate anomalous magnetic moment of the electron in quantum electrodynamics some mathematical trick, renormalisation, is used. It does not give mathematical self-consistent of quantum electrodynamics. In our case the theory is self-consistent.

Let us now consider how interaction of particles takes place in the model proposed. Let us suppose that a proton like an electron also forms the discrete fields. The model of a proton is much complicated. It will be described in further publications. Binding in atoms electron and proton also form discrete fields round themselves.

If two charges interact between each other by a field so this field is an object of matter and, therefore, should possess some elasticity. Namely the field not the vacuum is considered as an elastic medium round the particle. Equations of movement of an elastic medium can be described by well known equations for deviations of compression and shift [4]:

$$\frac{\partial^2 \mathbf{u}}{\partial t^2} - c_l^2\, graddiv\ \mathbf{u} + c_t^2\, rotrot\ \mathbf{u} = \mathbf{F} \tag{4}$$

where $\mathbf{u}$ is vector of deviation of the field, $\mathbf{F}$ is the vector of a force external to the medium, $c_t$ is the transversal wave speed and $c_l$ is the longitudinal wave speed for elastic medium. Maxwell and Newton equations are derived from this equation as partial cases. This equation, essentially, unites classical mechanics and electrodynamics.

Interaction between particles is realised by waves linked with an elastic medium and moving along planes crossing a tore.

For instance, according to our model boundary conditions for equation (4) applying to coupling electrons are defined by the form of the field like cross-section of a tore when electrons are at the opposite parts of such disc.

The theory of elasticity gives that internal strengthes coursed by the external forses act in compression-stretching like $1/R^2$ and for bending like $1/R^3$ [4]. This dependence of force on distance provides description between charges and magnetic dipoles. In other words, Coulomb's and Ampere's laws follow from the model naturally. In our model interaction is realised by longitudinal waves in matter but not by photons exchanging.

From the equation (4) is followed that the longitudinal speed is always more than transversal one, at least, in $2^{1/2}$ times. If two interacting particles move relative each other with the speed $v_0 \ll c$ so longitudinal wave speed will be much more than the speed of light if one take into account the relation $c_l c_t = c^2$. This condition was firstly formulated by de Broile in his dissertation. But he can not explain the mechanism of 'up-to-light' interactions. Later on his result was constantly ignored. Let us note that there was no experimental confirmation that speeds of interaction of slowly moving particles are equal to the speed of light. All interactions, as usual, are carried out at relativistic speeds.

It is automatically follows from the equation (4) that the transversal wave can not penetrate with the speed more than the speed of light that it was noted by Maxwell in his basic work. So no particle which is a transversal wave itself, and, particularly, condensed matter cannot move with speeds exceeding the speed of light. Interaction between particles by longitudinal waves can penetrate by speeds much exceeding the speed of light what can be used for transmission of information in the Galaxy scale. The problem of physical interpretation of longitudinal waves was met yet by Maxwell while he derived his equations. Later on these waves were omitted as having no physical sense.

On the base of the model proposed the most complicated effects were calculated. They are, first of all, points of phase transition of types I and II. The most important conclusions connected with phase transitions of the type II as macroquantum effects are given without deriving in table1.

Our model of electron and its movement in condensed media all deficiencies of existing models of superfluidity and superconductivity are eliminated automatically. In theory of superconductivity the retarding quantities connected with transversal and longitudinal wave speeds are introduced (3). Limiting parameters calculated according to the model proposed have high coincidence with experimental data for high temperature superconductivity without correcting coefficients. At that, our expressions include only the world constants and coefficients connected with the geometry of fields, boundary conditions and character of motion.

To verify the reliability of the model macroquantum effects at phase transitions in the processes of the stars and Solar system generation were considered [5]. Preliminary analysis allow to state that there is a phenomenon of high temperature superfluidity in atomic hydrogen at temperatures like $6000°C$.

The short summary of results is presented in the table 1.

Table 1

## *Summary of formulas for macroquantum effects in condensed matter*

| N | Content | Theoretical formula | Theoretical value | Experimental value | Ref |
|---|---|---|---|---|---|
| | | *This paper* | | *Independent experiment* | |
| | **Superfluidity** | | | | |
| 1. | Critical speed of superfluid phase motion relative to normal phase in a liquid $^4$HeII | $v_{max} = \dfrac{a^4 c}{\sqrt{2}}$ | 0.60011 m/s | 0.60 m/s | [1] |
| 2. | Limiting speed of the first sound in liquid helium | $v_1 = a^3 c \sqrt{\dfrac{4p}{3}}$ | 238.4303 m/s | 238.3 ± 0,1 m/s (pressure of saturated vapor at T=0,1K) | [1] |
| 3. | Limiting speed of the second sound in liquid helium | $v_2 = \dfrac{v_1}{\sqrt{3}} = a^3 c \dfrac{\sqrt{4p}}{3}$ | 137.58 m/s | 137.58 m/s (pressure of saturated vapor at T=0,1K) | [1] |
| 4. | Critical speed of sound in phase transition point | $v_\lambda = \dfrac{v_1}{\sqrt{2p}}$ | 95.12 m/s | at $T_\lambda$ | |
| 5. | Critical temperature of liquid $^4$He for phase transition in superfluid state | $T_\lambda = \dfrac{M v_\lambda^2}{2k} = \dfrac{M(a^3 c)^2}{3k}$ | 2.1780 K | 2.1720 K | [1] |
| | **Low temperature superconductivity** | | | | |
| 6. | Critical transition superconductivity temperature in pure metals | $T_c = \dfrac{M_i v_c^2}{2k}$ | | | |
| 7. | Minimal critical transition superconductivity temperature in pure metals | $T_c = \dfrac{M_i (a^4 c)^2}{4k}$ | 5.32·10⁻⁴ K | 5·10⁻⁴ K (Mg) | [11] |
| 8. | Maximum speed of sound in metals | $v_s = \dfrac{3a^2 c}{\sqrt{4p}}$ | 1.3509·10⁴ m/s | 1.341·10⁴ m/s (in Be along crystal axis $L_{001}$) | [12] |
| 9. | Critical speed of superconducting electrons in Be | $v_c = \dfrac{3a^3 c}{\sqrt{2p}}$ | 139.43 m/s | | [13] |
| 10. | Maximum critical transition superconductivity temperature in pure metals | $T_c = \dfrac{M_i}{2k}\left(\dfrac{3a^3 c}{\sqrt{2p}}\right)^2$ | 10.598 K | 10.5 K (in films of Be ≤3 nm), | [13] |
| | **High temperature superconductivity** | | | | |
| 11. | Critical temperature of high temperature superconductivity | $T_c = \dfrac{m_e^* v_c^2}{2k} = \dfrac{m_e (a^2 c)^2}{2p\, n a k}$ <br> $m_e^* = 2 m_e a^{-1}$, n = 1,2,3… | | | [8] |
| 12. | Maximum critical temperature for superconductors | n=1 | $T_{n1}$ = 366.65K (93.5°C) | 365K (92°C), powder-like superconductors on the base of XCuBr.CuBr$_2$ | [8, 14, 15, 16, 17] |
| 13. | | n=2 | n=2, $T_{n2}$=183.2K (-89.95°C) | 185K (-88°C) powders on the base of C$_{60}$/Cu in fraction of 7/1 | [18] |

| # | | | | | |
|---|---|---|---|---|---|
| 14. | | n=4 | $T_{c4}$=91.66K (-181.49$^0$C) | 91.6K (-181.55$^0$C) for superconductors like $YBa_2Cu_3O_7$ | [19] |
| 15. | | n=16 | $T_{c16}$=22.92K (-250.23$^0$C) | $T_ñ$= 23.2 ± 0.2K (-250.15$^0$C) 200 nm films of $Nb_3Ge$ | [11] |
| 16. | | n=32 | $T_{c32}$=11.46K (-261.69$^0$C) | Do not exceed in superconductors of type II: $Y_2C_3$, NbC, $Nb_3Au$ | [11] |
| 17. | Maximum frequency for electromagnetic waves in superconductors | $f_e = \alpha^2 c / 2\pi r_0 = m_e(\alpha^2 c)^2/h$ | 3.5037·10$^{11}$ Hz | | [8] |
| 18. | Limiting current density | $j_e = ef_e / (\pi r_0^2) = (4\pi e m_e^3 \alpha^8 c^4)/h^3$ | 3.4·10$^4$ A/cm$^2$ | | [8] |
| 19. | Critical magnetic field in high temperature superconductors | $B_e = 2\pi f_e m_e / e = (2\pi m_e^2 \alpha^4 c^2)/(eh)$ | 12.5 T | | [8] |
| 20. | Critical magnetic flow in high temperature superconductors | $\Phi_e = \pi r_0^2 B_e = h/2e$ | 2.0678·10$^{-15}$ Wb | | [8] |
| | **Solar system** | | | | |
| 21. | Critical orbit speed of movement in the Solar system | $v_1 = 3\alpha^2 c$ | 47.89307 km/s | 47.89 km/s equal to average orbit speed of the Mercury | [5, 20] |
| 22. | Critical speed of electrons relative to protons in the Solar shell | $v_{\Theta II} = \dfrac{\alpha c}{\sqrt{4\pi}}$ | 617.13 km/s | 617.7 km/s equal to the second cosmic speed for the Sun | [5, 20] |
| 23. | Temperature of electron gas in the Solar shell | $T_\Theta = \dfrac{m_e v_{\Theta I}^2}{2k}$ | 6282.1K | 6270.0 K equal to the temperature in the center of the Solar disc | [5, 20] |

$N_a$ = 861 is transversal quantum number; $\alpha$ is fine structure constant or longitudinal quantum number; e is elementary charge; $\hbar = \dfrac{h}{2\pi}$ is the Planck constant; $c$ is the speed of light; М is the mass of $^4$He atom; $M_i$ is the ion mass in the crystal lattice; $m_e$ is the mass of electron; $m_p$ is the mass of proton; k is the Boltzman constant; $m_e/\alpha$ is so called heavy mass of the electron.

The table shows that laws in macro and micro world are the same. One able to measure macroquantum effects very precisely because it is possible to use "three-dimensional" devises. In micro world "three-dimensional" devises reflect two-dimensional objects on thre-dimensional world. Because of segmentation of elementary particles the uncertainty like Heisenberg one is occurred. Probability description of elementary particles followed from this principle can be, therefore, overcome.

Analysis of data in table 1 gives that the picture of the world proposed also to describe the most complicated physical effects and to eliminate basic paradoxes appeared in the physics of the 20$^{th}$ century. There is no necessity to base on complex philosophical and mathematical concepts from which one can not obtain a mathematical approach appropriate to calculate experimental data. Our model yields to one of the basic criteria of the theory: it has a limit coincided with previous theories tested on the experiment.

**Acknowledgements**

The author acknowledge to the personnel of Consulting Center "Nanobiology" and Atomic and Molecular Engineering Laboratory for many years co-operation and support.